\documentclass[journal=chreay,manuscript=review]{achemso}

\usepackage[version=3]{mhchem} 
\usepackage{color}
\usepackage{enumerate}
\usepackage[normalem]{ulem}
\usepackage{hyperref}
\usepackage{lipsum}
\usepackage{setspace}

\hypersetup{
    colorlinks,
    citecolor=black,
    filecolor=black,
    linkcolor=black,
    urlcolor=black
}
\usepackage{float}

\def\bpf#1{\textcolor{blue}{#1}} 
\definecolor{ao(english)}{rgb}{0.0, 0.5, 0.0}

\author{Michael Woerner}
\email{woerner@mbi-berlin.de}
\author{Benjamin P. Fingerhut}
\email{fingerhut@mbi-berlin.de}

\author{Thomas Elsaesser}
\email{elsasser@mbi-berlin.de}
\affiliation[MBI]
{Max-Born-Institut f\"ur Nichtlineare Optik und Kurzzeitspektroskopie, D-12489 Berlin, Germany}

\title[]
  {Field-Induced Electron Generation in Water: Solvation Dynamics and Many-Body Interactions}


\abbreviations{2D-IR,DLS,DR,FFCF,H-bond,IR,MD,NMR,OKE,QENS,SAXS,SANS,tcf,TDSS,THz}

\begin{document}


\begin{abstract}
\footnotesize
The solvated electron represents an elementary quantum system in a liquid environment. Electrons solvated in water have raised strong interest because of their prototypical properties, their role in radiation chemistry, and their relevance for charge separation and transport.  Nonequilibrium dynamics of photogenerated electrons in water occur on ultrafast time scales and include charge transfer, localization, and energy dissipation processes. We present new insight in the role of fluctuating electric fields of the liquid for generating electrons in presence of an external terahertz field and address polaronic many-body properties of solvated electrons. The article combines a review of recent results from experiment and theory with a discussion of basic electric interactions of electrons in water.\\[2mm]

\end{abstract}


\newpage


\section{Introduction}

Electrons and protons solvated in liquid water are among the most elementary chemical species. They represent prototypic  quantum systems embedded in a fluctuating environment without long-range structural order. Interactions between the charged particles and their solvation shell include electric as well as other attractive or repulsive forces from noncovalent couplings. Vice versa, the presence of electrons and/or protons affects the local structure of the liquid and, upon nonequilibrium excitation, induces ultrafast solvation and energy dissipation processes. Such properties make solvated charges a direct probe of liquid dynamics. Beyond such basic aspects, solvated electrons and protons are relevant for the self-dissociation of water, charge transport in and  radiolysis of aqueous systems, e.g., of hydrated DNA and RNA structures. 

Solvated electrons have attracted strong and continued interest from both theory and experiment \cite{Rossky2012,Herbert2019,Marsalek2012,Alizadeh2012}. The so-called cavity model describes an electron in an excluded volume, surrounded by water molecules which form mainly electrostatic hydrogen bonds with the charged particle. The orientation of solvent molecules by interaction with the electron results in a reduction of free energy and, thus, a binding electron potential of a depth of some 5 eV. The wavefunction of the electron ground state is essentially localized within the cavity but displays some overlap with the first few water shells. 
This description has been challenged by a non-cavity picture according to which the electron's spin density is delocalized over several water molecules and induces an enhancement of water density near the center of the electron wavefunction.\cite{Larsen2010} 
Disputing this view, it was argued that the absence of a cavity was the result of a problem with the pseudo-potential parametrization.\cite{Jacobson:Science:2011,Turi:Science:2011,Larsen:Science:2011} Following a  vigorous debate, the cavity model is now widely accepted.\cite{Herbert:JPCA:2011,Uhlig:JPCL:2012,Casey:PNAS:2013,Casey:JPCB:2013,Ambrosio:JPCL:2017,Wilhelm:Angew:2019,Glover:JCTC:2020}

Multiphoton ionization of water by femtosecond pulses has allowed for mapping ultrafast nonequilibrium dynamics of solvated electrons in pump-probe and photon-echo experiments.\cite{Migus1987,Alfano1993,Emde1998,Laenen2000,Kambhampati2002,Savolainen2014}. The transition from initially delocalized electronic states to a localized bound quantum state in a self-consistent potential has been probed via the related spectral shift of optical absorption of the electron from the terahertz (THz) to the near-infrared spectral range. Localization occurs on a subpicosecond time scale, in parallel to the dissipation of excess energy which extends up to some 10 ps. After excitation of pre-existing solvated electrons from their electronic ground to higher electronic states, a sequence of subpicosecond relaxation processes between excited electronic states has been observed.

Recently, a novel generation mechanism of electrons based on tunneling ionization of water molecules in presence of an external THz field has been demonstrated \cite{Ghalgaoui2020}. The extremely high fluctuating electric field from water molecules in thermal motion spontaneously induces tunneling ionization of individual water molecules while irreversibility of the process is achieved via the separation of the electron from its parent ion in the THz field. The dynamics of such processes and the relevant molecular interactions are discussed in this article.    

The many-body character of Coulomb interactions between the electron and its fluctuating surrounding and between polar solvent molecules leads to local-field effects with a direct impact on the character of excitations of the solvated electron. As a consequence, nuclear and electronic degrees of freedom are being coupled and polaronic properties arise, beyond what is captured by existing one-electron models. Insight in this important aspect has remained very limited, in particular from the experimental side. Most recently, first experimental evidence for polaron excitations of solvated electrons has been presented \cite{Ghalgaoui2021}. 

In this article, we combine a short review of recent advances in understanding nonequilibrium dynamics of solvated electrons with a discussion of basic electric interactions of electrons in water. The role of fluctuating electric fields for generating electrons in an external THz field is addressed and new insight in Coulomb many-body interactions and polaronic properties of solvated electrons presented.
The content is organized as follows. Section 2 gives a summary of theoretical and experimental knowledge on structural properties and nonequilibrium dynamics of bulk water and solvated electrons. In Section 3, we discuss how the very high local electric fields in water induce tunneling ionization of water molecules and the separation of the electron from its parent ion in presence of an external THz field. Section 4 focuses on many-body aspects of Coulomb forces and addresses the polaronic properties of solvated electrons. Conclusions and an outlook are given in Section 5.

\section{Electrons solvated in water: Structure and Spectroscopy}
\label{sec:shell_def}

\subsection{Electric interactions and dynamics of bulk water}

Water molecules in the liquid phase display a permanent electric dipole moment of approximately 3 Debye which is substantially larger than  the dipole moment of an isolated water molecule of 1.86 Debye. This dipole enhancement is due to polarization by the electric field the other molecules in the liquid impose on a particular molecule \cite{Silvestrelli1999}. In addition to long-range Coulomb forces, local electrostatics result in the formation of intermolecular O-H...O hydrogen bonds.
At ambient temperature, water molecules undergo thermally activated motions on a femto- to picosecond time scale, including translations and librations. Their stochastic character leads to fluctuations in the arrangement of molecules in space, the hydrogen bond pattern, and  the nearest neighbor coordination number. The fluctuations give rise to a rapid  loss of intermolecular structural correlation.\cite{Bakker2009,Nibbering2004} Moreover, large angular jumps of water molecules induce hydrogen bond breaking on a time scale on the order of 1 ps \cite{Laage2006}. 

\begin{figure}[t]
\includegraphics[width=0.57\columnwidth]{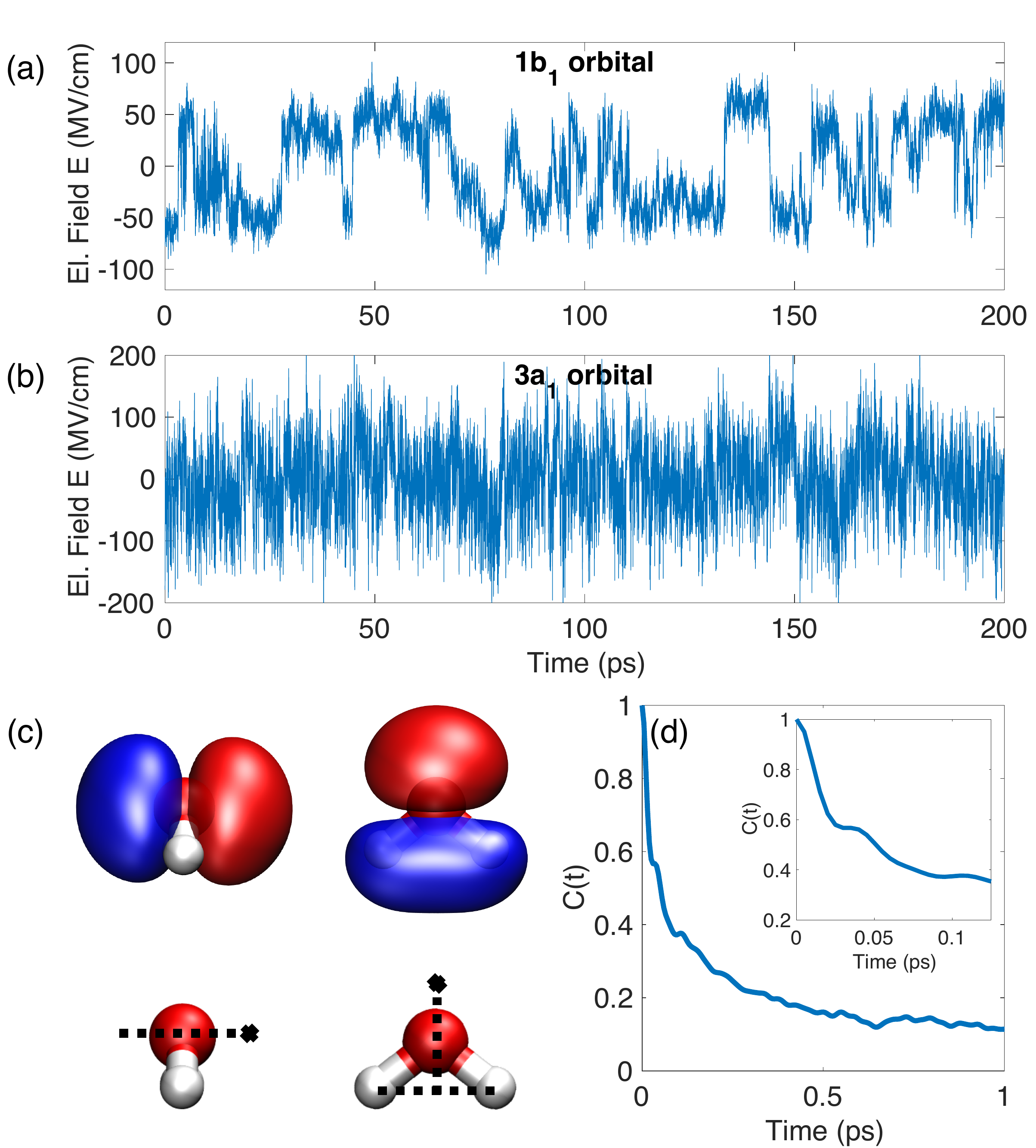}
\caption{(a)~Electric field trajectory calculated from a molecular dynamic (MD) simulation of neat water using the TIP4P-FB water model. 
MD simulations were performed for a cubic box of 216 water molecules with periodic boundary conditions (box size $\sim$ 1.87 nm) as described in detail in Ref.~\citenum{Ghalgaoui2020}.
The electric field projected on the axis of $1b_1$ molecular orbital (HOMO) of a water molecule shown in panel (c) is plotted as a function of time. (b) Same as (a) for an electric field projected on the vertical axis of the $3a_1$ orbital (HOMO-1) shown in panel (c). (c) Molecular $1b_1$ and $3a_1$ orbitals together with the projection axes (starred dashed lines) of the fluctuating electric field. (d) Time correlation function C(t) of the fluctuating electric field amplitude as derived from the MD simulations upon projection of the electric field vector on the axis of the $3a_1$ orbital [cf. Fig. 1(c)].}
\label{fig:MD}
\end{figure}

The stochastic molecular motions result in ultrafast fluctuations of the strength and direction of local electric fields. Molecular dynamics (MD) simulations of water have been performed with models of different sophistication, in particular with and without including flexible and polarizable water molecules.\cite{Vega2011,Lambros2020} Nevertheless, most MD simulations agree in predicting local electric field strengths of up to some 200 MV/cm with fluctuation amplitudes of several tens of MV/cm. In Fig. \ref{fig:MD}, electric field trajectories from MD simulations based on the TIP4P-FB water model are presented, covering a time range of 200 ps \cite{Ghalgaoui2020}. Panel (a) shows the electric field projected on the axis of the $1b_1$ molecular orbital (HOMO) of a water `probe' molecule (Fig. \ref{fig:MD}c). The electric field amplitudes are in a range of $\pm 100$ MV/cm. Even stronger fields arise along the axis of the $3a_1$ orbital [HOMO-1, panel (b)], where electric field spikes of an amplitude around 200 MV/cm and a sub-50 fs duration occur every 50 to 100 ps. At such instants in time, the probe molecule forms three hydrogen bonds only, two as a hydrogen donor and one as a hydrogen acceptor. The latter H-bond between the oxygen atom and an O-H group of a neighboring water molecule is particularly short with an O...O distance of 2.54 \AA~ and oriented parallel to the axis of the $3a_1$ orbital. 

In Fig. \ref{fig:MD}(d), the time correlation function $C(t) = \langle \delta E(t) \delta E(0) \rangle / \langle \delta E(0) \delta E(0) \rangle$ of the fluctuating electric field is shown, where $E(t)$ stands for the amplitude of the electric field projected on the axis of the $3a_1$ orbital [Fig. \ref{fig:MD}(c)] and $\delta E(t)$ for its excursion from the average value $\bar{E}$. The brackets represent the ensemble average. The correlation function  exhibits an initial decay on a 50-fs time scale and slower subpicosecond decay components. The initial decay is governed by thermally excited librations in the frequency range between 500 and 1000 cm$^{-1}$.\cite{cowan2005,Jansen2010} The fluctuating electric field acting on a polarizable water molecule distorts its electronic structure and leads to time-dependent changes of electronic and vibrational transition frequencies, i.e., spectral diffusion and decoherence of quantum-coherent excitations and the resulting macroscopic polarization. 

The combination of two-dimensional infrared (2D-IR) spectroscopy with theory and MD simulations has established a  quantitative picture of the ultrafast dynamics of neat water, aqueous solutions of ions, and hydrated biomolecules.\cite{Bakker2009,Nibbering2004,Laage2017}. Spectral diffusion has been followed in time via changes of 2D-IR line shapes of O-H stretching excitations which are governed by the frequency fluctuation correlation function (FFCF) $\langle \delta \nu (t) \delta \nu (0) \rangle$ of the transition frequency $\nu$.\cite{Mukamel2000} In turn, the FFCF is determined by the correlation function of the fluctuating electric field $C(t)$ and, thus, the underlying structural dynamics of the liquid. A link between the momentary electric field amplitude $E(t)$ and $\delta \nu (t)$ is established with the help of vibrational frequency maps $\nu(E)$. 

The FFCF of neat water exhibits an initial decay on a 50-fs time scale, due to high-frequency librational motions and followed by a decay on a time scale of several hundred femtoseconds. The latter is caused by molecular motions at lower frequency and, for O-H stretching excitations, resonant energy transfer between water molecules. In an even longer picosecond time range, hydrogen bond breaking and reformation contribute as well. In case the water structure is subject to particular steric boundary conditions, such as in the solvation shell of ions or at the surface of a biomolecule, a moderate lengthening of the shortest correlation decay arises \cite{Laage2017}. The vibrational lifetimes of O-H stretching and bending excitations in neat water are 200 fs and 170 fs, librational excitations decay on a sub-100 fs time scale.

\subsection{Theoretical description of the hydrated electron: the canonical cavity model}

\begin{figure}[t]
\includegraphics[width=0.52\columnwidth]{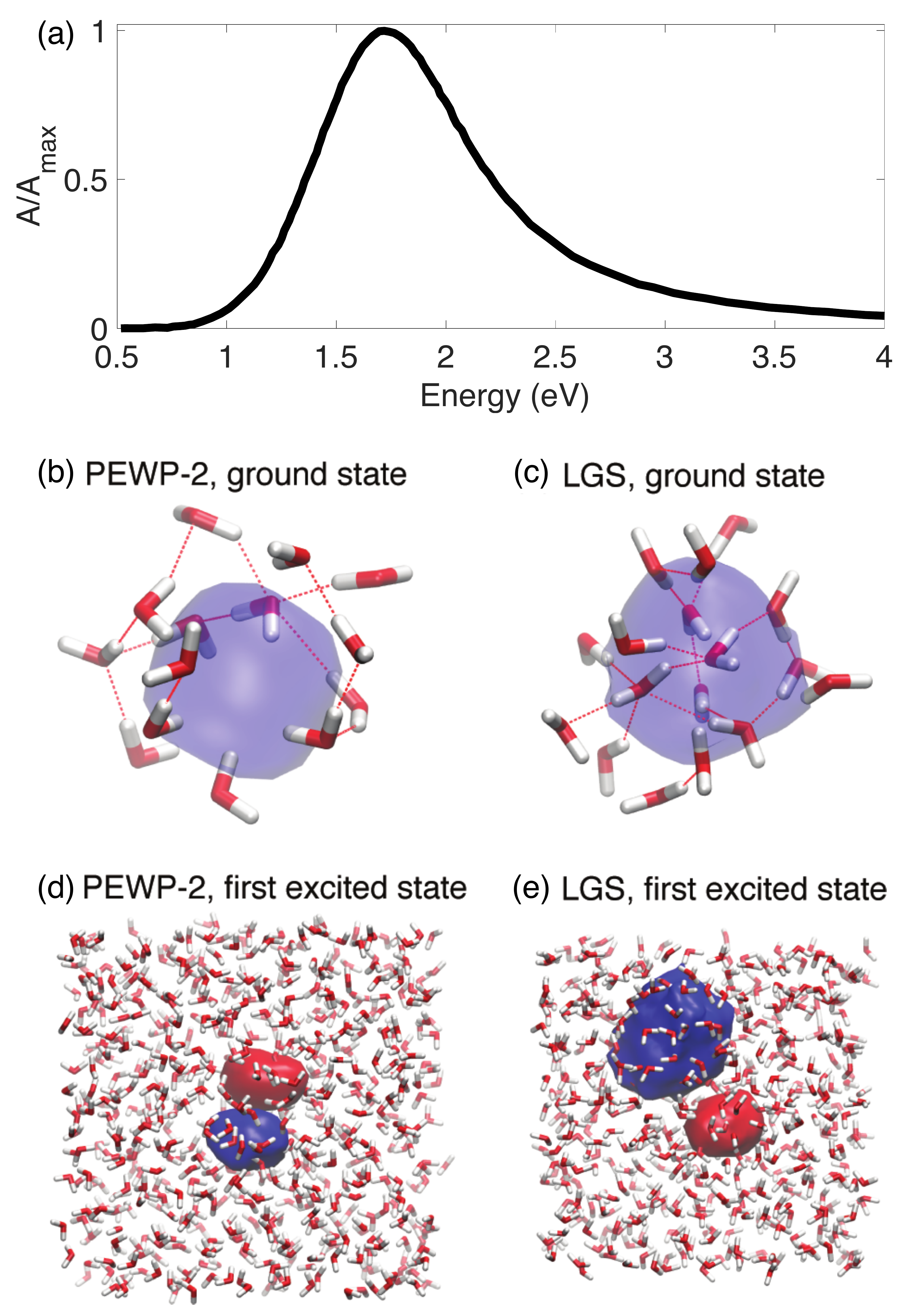}
\caption{
(a) Experimental optical absorption spectrum of the hydrated electron in water at T = 300 K.\cite{Jou:JPC:1977,Jou:CandJChem:1979}
(b) Ground state and (d) first excited state wavefunctions of the hydrated electron in the cavity forming model PEPW-2.
(c) Ground state and (e) first excited state wavefunctions of the hydrated electron in the non-cavity forming Larsen-Glover-Schwartz (LGS) model.
Panels (b-e) reprinted with permission from ref. ~\citenum{Herbert:JPCA:2011}. Copyright 2011 American Chemical Society.
}
\label{fig:CavityModel}
\end{figure}

\begin{figure}[t]
\includegraphics[width=0.45\columnwidth]{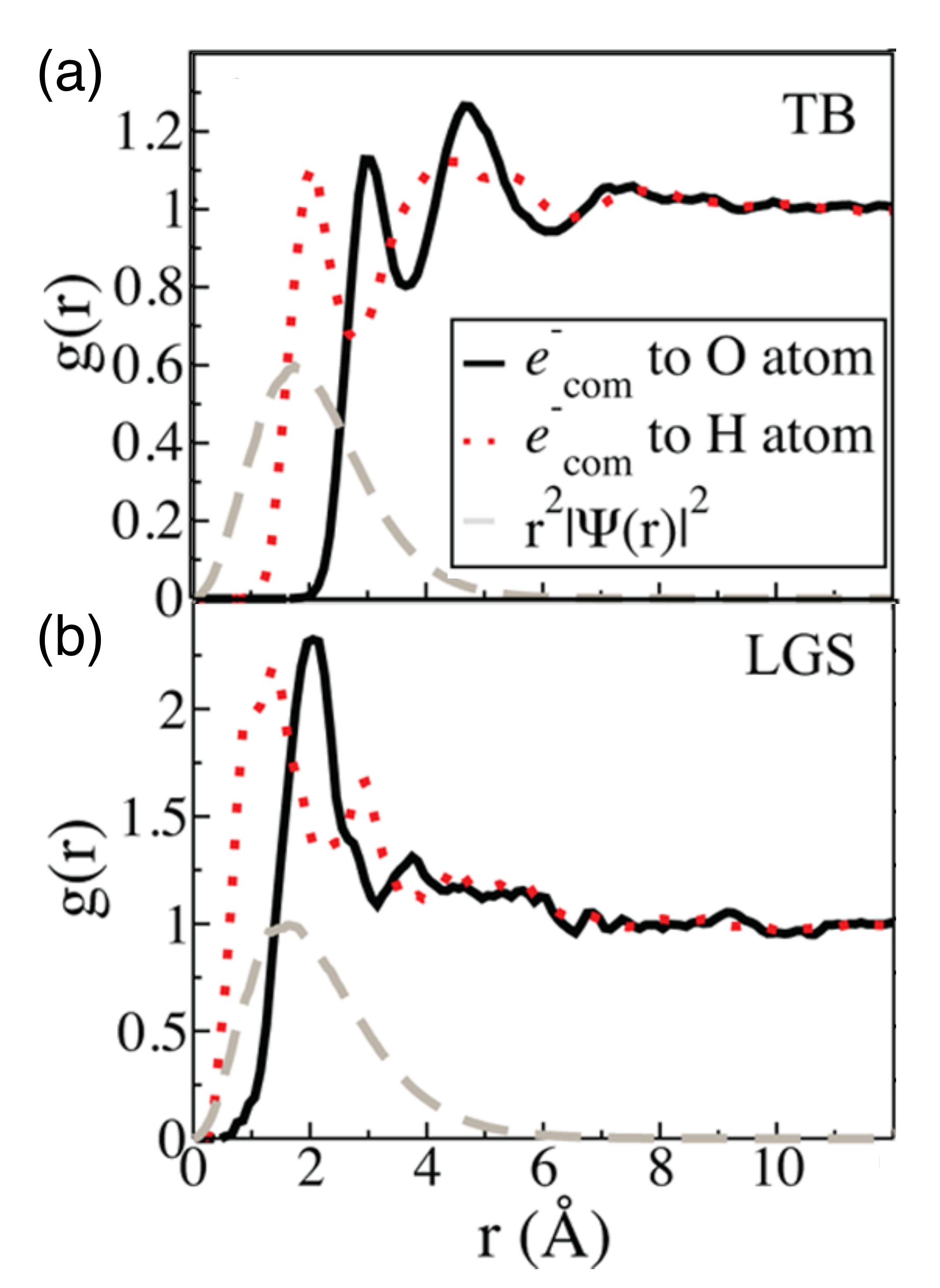}
\caption{
Electron center of mass - oxygen radial distribution functions (black, solid lines), electron center of mass - hydrogen radial distribution functions (red, dashed lines)
and electron density (gray, dashed lines) of (a) the cavity forming Turi-Borgis
 (TB)   model  and (b) the non-cavity forming Larsen-Glover-Schwartz (LGS) model.
Reprinted with permission from ref. ~\citenum{Casey:JPCB:2013}. Copyright 2013 American Chemical Society.
}
\label{fig:rdf}
\end{figure}

The molecular level understanding of the properties of the hydrated electron has been covered in recent comprehensive reviews, e.g., Refs.~\citenum{Rossky2012,Herbert2019}, while the basic structural motif relevant for hydrated electron stabilization by surrounding water molecules was outlined  already in 1953:\cite{Platzman:1953}
the trapping of the  electron in a potential well arises from a polarization of the surrounding water molecules via their orientation  under the influence of the negative charge (Fig. 2).
The thermalized electron thus occupies an excluded volume in the structure of liquid water and coordination of water molecules 
occurs via the directed, hydrogen bond-like interaction with a single hydroxyl group of water molecules in the first solvation shell.
At room temperature, the  instantaneous coordination number is fluctuating and the hydration structure notably deviates from an idealized Kevan structure which consists of an octahedral arrangement of six water molecules with one OH group each pointing to the electron.\cite{Feng:ChemRev:1980} 
The predicted average number of hydrogen bonds varies between 4 and 6, depending on the particular theoretical model and the employed  level of theory. 
Accordingly, the solvation pattern of the hydrated electron in the canonical cavity model\cite{Schnitker:JCP:1987,Turi:JCP:2002} resembles that of a pseudo-halide Fig.~\ref{fig:CavityModel}b,d). 
%
Different theoretical methods, ranging from single active electron quantum mechanical - molecular mechanical (QM/MM)
methods in the  Turi-Borgis (TB) model,\cite{Turi:JCP:2002} density functional theory (DFT),\cite{Ambrosio:JPCL:2017} 
and many body perturbation theory (MP2),\cite{Wilhelm:Angew:2019} predict the hydration structure in the first solvation shell (Fig.~\ref{fig:rdf}a) with high consistency.
For all methods, a characteristic of the cavity model is the breaking of water-water hydrogen bonds in the solvation shell surrounding the hydrated electron.

The formation of the cavity by a reorganization of the first few water shells leads to a stabilization of the electron ground state in a trap potential, characterized by an electron binding energy of $\approx$ 3.7 eV.\cite{David:SciAdv:2019}
The electronic transitions responsible for the prominent absorption band of the  hydrated electron (Fig.~\ref{fig:CavityModel}a) have been assigned to dipole-allowed  $s\rightarrow p$ excitations where fluctuations of the cavity potential due to solvent motions induce a breaking of the three-fold degeneracy. The wave function of the hydrated electron partially extends beyond the cavity (40–60 \% of the spin density are confined within the cavity) and stretches out approximately up to two water shells (Fig.~\ref{fig:rdf}a).\cite{Uhlig:JPCL:2012}

The cavity model has been  invoked to explain a range of properties of hydrated electrons and accounts for a large portion of the broad optical absorption spectrum (see Sec.~\ref{sec:ExpSimulations}). 
The radius of gyration,  the root-mean-square distance of electronic charge from its center of gravity, is frequently used to quantify the spatial extent of the electron wave function. It shows a strong inverse correlation with the electronic excitation energy.\cite{Herbert2019}
Such correlation of the electronic excitation energy  and gyration radius provides access to the electron localization dynamics following generation of nonequilibrium excess electrons. The transition from nonequilibrium delocalized  states to trap states confined in the cavity is connected with a shrinking radius of gyration and a concomitant shift of electronic absorption from the THz to the NIR spectral region,\cite{Savolainen2014} in qualitative agreement with the estimate of the initial spatial wavefunction delocalization over some 40 \AA. 

An alternative view to the cavity model was proposed where the electron occupies a region of enhanced water density of some $\sim$10 \AA~diameter (Fig.~\ref{fig:CavityModel}c,e).\cite{Larsen2010}
Within the first minimum of the electron-oxygen radial distribution function, $\approx$37 water molecules are contained, compared to only $\sim$30 water molecules expected for neat water, corresponding to a locally increased water density of $\rho \approx 1.2$ g/cm$^3$ within the spatial extent of the electron.\cite{Casey:JPCB:2013} 
In this so-called Larsen-Glover-Schwartz (LGS) model a cavity is absent and the electron has a rather delocalized character.
Within the electron volume, a net orientation of water molecules is induced with a preference of the water O-H bonds pointing to the center of charge of the hydrated electron, but the hydrogen bond network within this volume is largely preserved. 
The charge density of the electron shows appreciable spatial overlap with the density of water molecules and the hydrated electron extends beyond the first hydrogen and oxygen peaks of the radial distribution function.
The region of enhanced water density extends to some $\sim$ 6 \AA~from where on bulk behavior is observed (Fig.~\ref{fig:rdf}b).
 Almost immediately after publication of the non-cavity model criticism of the employed pseudo-potential parametrization was raised\cite{Jacobson:Science:2011,Turi:Science:2011,Larsen:Science:2011} (see also Refs.~\citenum{Herbert:JPCA:2011,Uhlig:JPCL:2012,Casey:PNAS:2013,Casey:JPCB:2013,Herbert2019,Dasgupta:JPCB:2019} on the cavity vs. non-cavity model debate).
 Recently,  short-range correlation effects were suggested to be particularly important for stabilizing non-cavity solvation structures of the condensed-phase hydrated electron.\cite{Glover:JCTC:2016,Glover:JCTC:2020}
Nevertheless, due to the largely consistent predictions of  the electron hydration structure  from ab-initio and DFT simulations\cite{Ambrosio:JPCL:2017,Wilhelm:Angew:2019}  and the TB single-active electron model,
today the cavity model with the breaking of water-water hydrogen bonds in the solvation shell surrounding the hydrated electron is the widely accepted picture.

Such ab initio and first-principles approaches to hydrated electrons\cite{Boero:PRL:2003,Uhlig:JPCL:2012,Ambrosio:JPCL:2017,Pizzochero:ChemSci:2019,Wilhelm:Angew:2019}  mitigate shortcomings of single-active-electron approaches by inherently accounting for many-body polarization and intermolecular charge transfer. 
In particular, recent molecular dynamics simulations relying on hybrid density functionals\cite{Ambrosio:JPCL:2017} have reinforced the cavity view of the hydrated electron and provide an accurate description of the band gap of liquid water and the electron. 
The hydrated electron was found to localize in a cavity with coordination numbers fluctuating  in the range 4 - 6 and the reported cavities were found to be slightly more compact  than in the single active electron TB model.   The radius of gyration (2.49 \AA) as well as the calculated peak position of 1.75 eV of the optical absorption spectrum agree well with experimental values. 
Non-equilibrium 
cavity formation involves the reorganization of the hydrogen-bond network on a few-hundred femtosecond time scale,\cite{Pizzochero:ChemSci:2019}   as recently corroborated 
by molecular dynamics simulations relying on many-body perturbation theory.\cite{Wilhelm:Angew:2019} 

\subsection{Relation of the hydrated electron structure to experimental observables}
\label{sec:ExpSimulations}

\subsubsection{Optical absorption spectrum}

The optical absorption spectrum in the cavity model arises from dipole-allowed $s\rightarrow p$ transitions between localized electron states (Fig.~\ref{fig:CavityModel}d), being energetically separated from (dark) transitions to continuum states.
Despite the success of the TB single-active electron model to accurately describe the peak position of the optical absorption of the hydrated electron, the model  fails to correctly reproduce the characteristic asymmetric line shape.\cite{Hart:JACS:1962,Jou:JPC:1977,Jou:CandJChem:1979} 
Substantially improved agreement of the cavity model with the experimental absorption spectrum (Fig.~\ref{fig:CavityModel}a) is obtained when including polarization and nuclear quantum effects.
Mutual many-body polarization between the hydrated electron and surrounding solvent  induces a $\sim$0.3 eV red-shift of the absorption maximum 
 and leads to a tail on the blue edge of the simulated spectrum due to intensity borrowing by quasi-continuum excited states.  \cite{Jacobson:JCP:2010,Jacobson:JACS:2010,Herbert:IntRevPhysChem:2011}
 Inclusion of solvent electronic polarizability and electronic relaxation of the solvent upon excitation of the one-electron wave function thus fine tunes 
 the energetic position of the band maximum, allowing for almost quantitative agreement to the experiment.
Nuclear quantum effects considered in electronically adiabatic quantum time correlation function simulations\cite{Turi:JCP:2009} lead to 
an improved line shape  on the low energy side of the spectrum and increased intensity on the high energy side of the optical absorption spectrum, however, insufficient to resolve persistent deviations in the Lorentzian-shaped  part of the spectrum in the 2–3 eV region.
The fact that both nuclear quantum effects and many body polarization allow for improving  the agreement  to the experiment reflects the inherent many-body effects in the interaction of the  hydrated electron with the condensed phase.

\subsubsection{Femtosecond solvation dynamics}

The line shape of the electronic absorption spectrum of the hydrated electron and the underlying broadening mechanisms are a matter of a  longstanding debate.
Polarized pump-probe hole-burning experiments were interpreted in terms of a predominant homogeneous broadening.\cite{Cavanagh:CPL:2004}
Such findings are in line with reports of a vanishing anisotropy on the ultrafast time scale and the absence of spectral hole burning around the excitation wavelength.\cite{Assel:JPCA:1998,Assel:CPL:2000} 
Both observations suggest a  fast population redistribution among $p$ states and/or fast (sub-100 fs) solvation dynamics.
Photon-echo peak shift measurements  with a sub-20-fs time resolution\cite{Emde1998} reveal a strong coupling of optical transitions to $\sim$850 cm$^{-1}$ librations, i.e., hindered rotations around the axis parallel to the H-H connecting line in an H$_2$O molecule.

In the cavity model, fluctuations of $s$ ground state and $p$ excited state energies are induced by 
solvent-induced distortions of the spherical cavity structure and  a certain degree of inhomogeneous broadening is thus expected.
On top, many-body polarization has been suggested as important contribution  to explain 
the transient hole burning experiments.\cite{Jacobson:JCP:2010}
Due to the electronic relaxation of the solvent upon electronic excitation, 
the overlap of the different $s \rightarrow p$ sub-bands with quasicontinuum transitions is increased, leading to a redistribution of dipole transition strength.
The Fourier transforms of the energy gap and dipole correlation function further indicate that  translational and librational modes dominate the solvent response subject to excitation of the electronic degrees of freedom,\cite{Turi:JCP:2009} qualitatively consistent with the results from photon-echo peak shift measurements.\cite{Emde1998}

\subsubsection{Resonance Raman spectrum of the hydrated electron}

The resonance Raman spectra of the hydrated electron\cite{Mizuno:JPCA:2001,Tauber:CPL:2002,Tauber:JACS:2003}  suggest
a predominant homogenous broadening of the optical absorption.\cite{Tauber:CPL:2002}
They further suggest that water vibrations, 
presumably of molecules in direct interaction with the hydrated electron, are resonantly enhanced upon electronic excitation of the hydrated electron and that the stretching, bending and librational modes of water associated with the hydrated electron are red-shifted compared to bulk water. The strong Raman enhancement of librational bands indicates  that hindered rotations are a central component of the solvent response following $s \rightarrow p$ electronic excitation. \cite{Tauber:CPL:2002} 

Vibronic interactions  of electronic and vibrational degrees of freedom were revealed in picosecond time-resolved resonance Raman spectra.\cite{Mizuno:JPCA:2003}
The observations support the view of a quasi-molecule, composed 
of the electron and the surrounding water shell. This is in line with the femtosecond infrared response of the hydrated electron\cite{Thaller:CPL:2004} which displays short lived absorption features in the mid-infrared, centered around the water O-H/O-D stretching absorption band.  A pronounced coupling of electronic transitions of the solvated electron with the O-H groups of water molecules can be mediated through directed hydrogen bonds of water O-H groups to the electron.\cite{Tauber:JACS:2003}

A red shift of water O–H stretching vibrations, as observed in resonance Raman spectra  of hydrated electrons,\cite{Tauber:CPL:2002} 
is commonly  associated with an increased hydrogen bond strength and consistent with the red shift of the stretching mode  of high-density water under pressure.\cite{Corcelli:JCP:2004,Walrafen:JSolChem:1973}
 All-electron ab initio calculations suggest that  charge transfer from the unpaired electron into anti-bonding $\sigma^*$ orbitals of the O-H bond of hydrating water molecules induces the red-shift of O-H stretching vibrations.

In summary, detailed modeling of the structural properties and spectral observables of the hydrated electron in water 
reveals a subtle interaction pattern of many-body polarization, charge transfer and nuclear quantum effects that are important for a faithful agreement with experimental observations.
Vibronic coupling of the electronic degree\bpf{s}  of freedom of the hydrated electron and the vibrational degrees of freedom of the water environment 
suggests a view of a quasi-molecule composed of the hydrated electron and the first few water shells. 
Such findings pose challenges to simulations relying on single-active electron approximations.

The quasi-molecular view of strongly coupled electronic and nuclear degrees of freedom has been addressed in early polaron models that describe the hydrated electron as a spherical charge distribution with a radius determined self-consistently by polarization of the surrounding solvent.\cite{Weiss:Nature:1960,Jortner:1964,Marcus:JCP:1965,Webster:NatPhysSci:1972,Laria:JCP:1991}
Our very recent findings of long-lived polaron oscillations demonstrate that the dynamics of solvated electrons in the low-frequency THz spectral region is dominated by the many-body interactions of collective and strongly coupled electronic and nuclear degrees of freedom (cf. Sec.~\ref{sec:vib_exc_dissipation}).

\section{Electrons as a probe of fluctuating electric fields
\label{sec:Struct_dyn}
}

 \subsection{Field-induced ionization and charge separation}   

\begin{figure}[t]
\includegraphics[width=0.65\columnwidth]{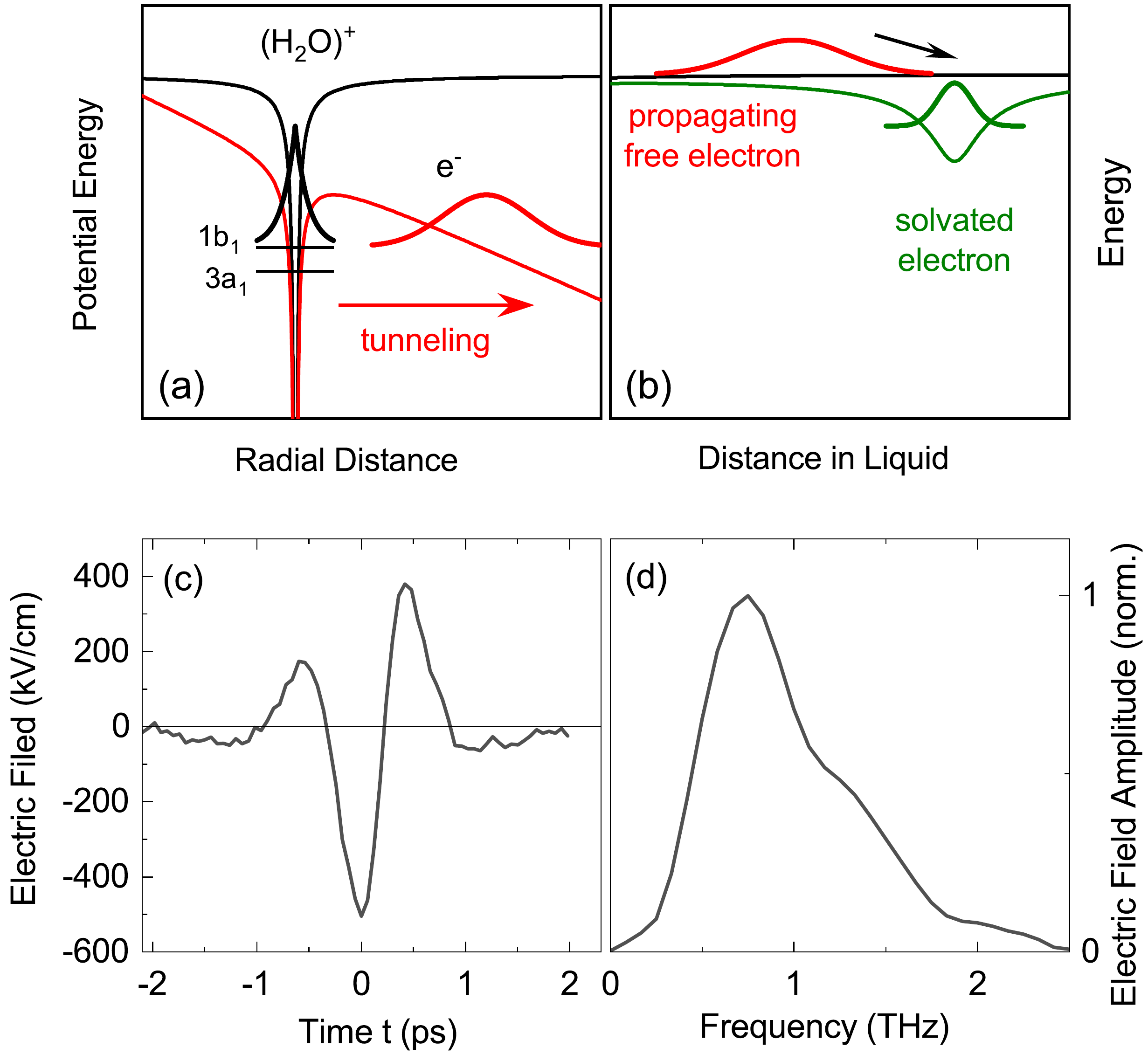}
\caption{(a)~Schematic of tunneling ionization of a water molecule in a strong external electric field. The distortion of the molecular potential by the electric field generates a tunneling barrier through which electrons from the highest $1b_1$ (HOMO) and second highest $3a_1$ (HOMO-1) molecular orbital can tunnel into continuum states. (b) Schematic of electron transport in an external THz field. The electron wavepacket propagates along the direction of the THz field and eventually localizes in a potential created by a solvation process. (c) THz pulse with a peak electric field of approximately 500 kV/cm. The electric field is plotted as a function of time. (d) Amplitude spectrum of the THz pulse in panel (c) with a maximum at 0.7 THz. Adapted from ref.~\citenum{Ghalgaoui2020} published under an American Chemical Society Author Choice License.}
\label{fig:tunnel}
\end{figure}

The highest sub-50 fs peaks of the fluctuating electric field in bulk water reach an amplitude on the order of 200 MV/cm (cf. Fig.~\ref{fig:MD}), which is sufficient for ionizing a water molecule by electron tunneling. The basic scenario is sketched in Fig. \ref{fig:tunnel}(a,b). The external electric field distorts the electronic potential of the molecule in a nonperturbative way, thus creating a tunneling barrier of finite width and height for electrons in the highest occupied molecular orbitals 1$b_1$ (HOMO) and 3$a_1$ (HOMO-1). Electrons can tunnel from the bound orbital states into continuum states outside the barrier, leaving an H$_2$O$^+$ parent ion behind. The released electron represents a wavepacket made up of a superposition of continuum states and propagating along the spatial electric field gradient. Upon propagation, the wavepacket spreads in space and, in parallel, loses its quantum-coherent character under the action of the fluctuating electric field from the water environment. As a result of such decoherence, propagation comes to an end and the electron localizes at a distance from the parent ion. For water in thermal equilibrium, this charge separation is not persistent but recombination of electron and parent ion, induced again by the fluctuating electric field, eventually prevails. 

In contrast to tunneling ionization of water molecules in the gas phase \cite{Farrell2011,Petretti2013}, this scenario lacks a detailed theoretical description and/or simulation. A first basic analysis has been presented in Ref. \citenum{Ghalgaoui2020} and combines the following elements: 
\begin{itemize}
\item{Electron tunneling and the very early phase of wavepacket propagation have been accounted for by solving the time-dependent Schrödinger equation for a one-dimensional rectangular potential subject to a 10-fs Gaussian electric field transient. The latter accounts for a short spike of the fluctuating electric field in the liquid. The tunneling probability has a value around $5 \times 10^{-12}$ for a peak electric field $E_{peak}=150$~MV/cm and rises to $10^{-6}$ for $E_{peak} \approx 300$~MV/cm. Averaging over the 200-ps period of the MD trajectory in Fig.~\ref{fig:MD}(b) gives an average tunneling probability of $3.4 \times 10^{-7}$ from the $3a_1$ orbital. The tunneling probability from the $1b_1$ orbital is much lower because of the smaller projected electric field (Fig.~\ref{fig:MD}a). }
\item{The wavepacket spreads to a several 10-\AA~width within the first few femtoseconds of propagation and reaches a distance of some 40 \AA~from the parent ion after 2-3 fs propagation time. It is important to note that the model describes an electron wavepacket which propagates ballistically in an external electric field. In real liquid water, this field is screened  on a length scale comparable to the estimated propagation length and additional random electric forces arise from the dipolar water molecules. Thus, the model  only covers the initial few femtoseconds of transport under the action of the electric-field spike. Propagation distances on a longer time scale can be derived from an analysis of experimental results and will be discussed in section 3.2.} 
\item{ Using the Caldeira-Leggett path integral approach \cite{Caldeira1983}, one estimates a decoherence rate $\Gamma_{decoh}  =2.6 \times 10^{14}$ ~s$^{-1}  \approx 5 \Gamma_{fluc}$ where $\Gamma_{fluc}=1/(20 \rm{fs}) =5 \times 10^{13}$~s$^{-1}$ is the fluctuation rate of the electric field in the liquid as derived from the fast decay of the field correlation function (Fig.~\ref{fig:MD}d).}
\item{The recombination process is described as a radiationless decay from the charge-separated state to the ground state of the parent molecule, induced by the fluctuating electric field. Quantum-classical MD simulations of radiationless processes of solvated electrons suggest a recombination time of 50-100 fs.\cite{Prezhdo1997}
In thermal equilibrium, the balance of electron generation and recombination leads to a steady-state concentration of free electron of several $10^{-7}$~M. }
\end{itemize} 
We note that charge separation and recombination are completed on a time scale much shorter than the picosecond lifetimes of intermolecular hydrogen bonds. In other words, there are no major changes in molecular arrangement during this period.

A persistent separation of the electron from the parent ion requires a suppression of the ultrafast recombination process. This is possible by applying an external directed electric field of sufficient strength along which the electron can  move a sufficient distance from the parent ion. For a persistent separation of charge, the electron has to acquire a minimum kinetic or ponderomotive energy $U_p$ [cf. eq. (1)] given by $U_p =  U_I \approx 11$~eV, where $U_I$ is the ionization potential of water in the liquid phase. 
 Electron transport is subject to frictional electric forces from the water environment, limiting the electron mobility. After switching off the external field, the electron dissipates its excess energy into the liquid and solvates in a new environment. 

A key issue in implementing this concept is the choice of the external directed electric field. The impact of the field on water structure and its fluctuations should be as small as possible in order to preserve the intrinsic properties of the liquid. This condition rules out static external fields which result in a spatial ordering of water dipoles and, for amplitudes in the MV/cm range, induce crystallization of supercooled water \cite{Svishchev1994,Svishchev1996}. Moreover, the DC mobility of electrons in water is comparably small \cite{Barnett1990}. Given the picosecond time scale of hydrogen bond breaking and reformation which represent spontaneous structure changes, electric field transients of a duration on the order of 1 ps appear appropriate. This time range corresponds to a carrier frequency of the electric field around 1 THz. 

The strength of the local directed electric field needs to be sufficient to accelerate the released electron to a ponderomotive energy $U_p = U_I$. $U_p$ is given by
\begin{equation}
U_p=\frac{e^2}{8 \pi^2 m_e \nu^2} \cdot \left[ \frac{\epsilon(\nu) + 2}{3} \right]^2  \cdot \vert E(\nu) \vert^2
\end{equation}
with the elementary charge $e$, the electron mass $m_e$, the dielectric constant of water $\epsilon(\nu)$, and the frequency $\nu$ and amplitude $E(\nu)$ of the external electric field. The second term on the right-hand side is the Clausius-Mosotti factor accounting for the local enhancement of the external field $E(\nu)$ in the liquid. For a carrier frequency $\nu = 1$~THz, a value of $U_{p} = U_I \approx 11$~eV is reached with an external field amplitude $E \approx 250$~kV/cm. Such field strengths are well in the range accessible with current THz generation technologies. In Figs.~\ref{fig:tunnel}(c,d), the time dependent electric field of a THz transient with a peak field of 500 kV/cm and its frequency spectrum with maximum at 0.7 THz are shown. 

Strong-field ionization and field-driven electron transport play a central role also in high-harmonic generation (HHG) which has recently been demonstrated in water and selected alcohols.\cite{Luu2018}. The external optical field inducing ionization as the initial step of HHG has a peak amplitude on the order of 100 MV/cm, i.e., similar to the maximum amplitude of the fluctuating electric field that ionizes water molecules spontaneously. In HHG, electron motion and the eventual recombination of the electron with the parent ion, the prerequisite for the emission of high harmonics, are steered by the directed coherent optical field and occur within a half cycle, i.e., a time scale of a few femtoseconds. In contrast, the recombination electron and parent ion without  optical field is induced by the fluctuating electric field in the liquid and has the character of an incoherent radiationless process. 

\subsection{Experimental results}

Tunneling ionization of water molecules in presence of a strong THz electric field has recently been studied in ultrafast THz pump/ optical probe experiments and by 2D-THz spectroscopy \cite{Ghalgaoui2020,Reimann2021}. Single-cycle THz pulses with an electric field amplitude up to 2 MV/cm and a 1-kHz repetition rate were generated by tilted-wavefront optical rectification of near-infrared pulses in a LiNbO$_3$ crystal.\cite{Hebling2008}. The THz transients were detected by phase-resolving free-space electrooptic sampling in a 10-$\mu$m thick ZnTe crystal.\cite{Reimann2021} This detection method gives the THz electric field as a function of real time $t$, as shown in Fig.~\ref{fig:tunnel}(c). A femtosecond white-light continuum in the visible/near-infrared spectral range served for probing changes of electronic absorption of the water sample induced by THz excitation. 

For the 2D-THz experiment, a phase-locked pair of THz pulses A and B separated by the delay time $\tau$ interacted with the sample in a transmission geometry. The time dependent nonlinear signal field $E_{NL}(t,\tau)$ is given by $E_{NL}(t,\tau) = E_{AB}(t,\tau) - E_A(t,\tau) - E_B(t)$, where $E_{AB}(t,\tau)$ is the electric field transmitted through the sample after interaction with two THz pulses A and B, and $E_A(t,\tau)$ and $E_B(t)$ are the transmitted fields after interaction with pulse A or B only \cite{Reimann2021,Elsaesser2019}.  A double Fourier transform of  $E_{NL}(t,\tau)$ along $t$ and $\tau$ provides the nonlinear signal as a function of excitation frequency $\nu_\tau$ and detection frequency $\nu_t$. The water sample was a free-flowing jet of 50 $\mu$m thickness. 

\begin{figure}[t]
\includegraphics[width=0.44\columnwidth]{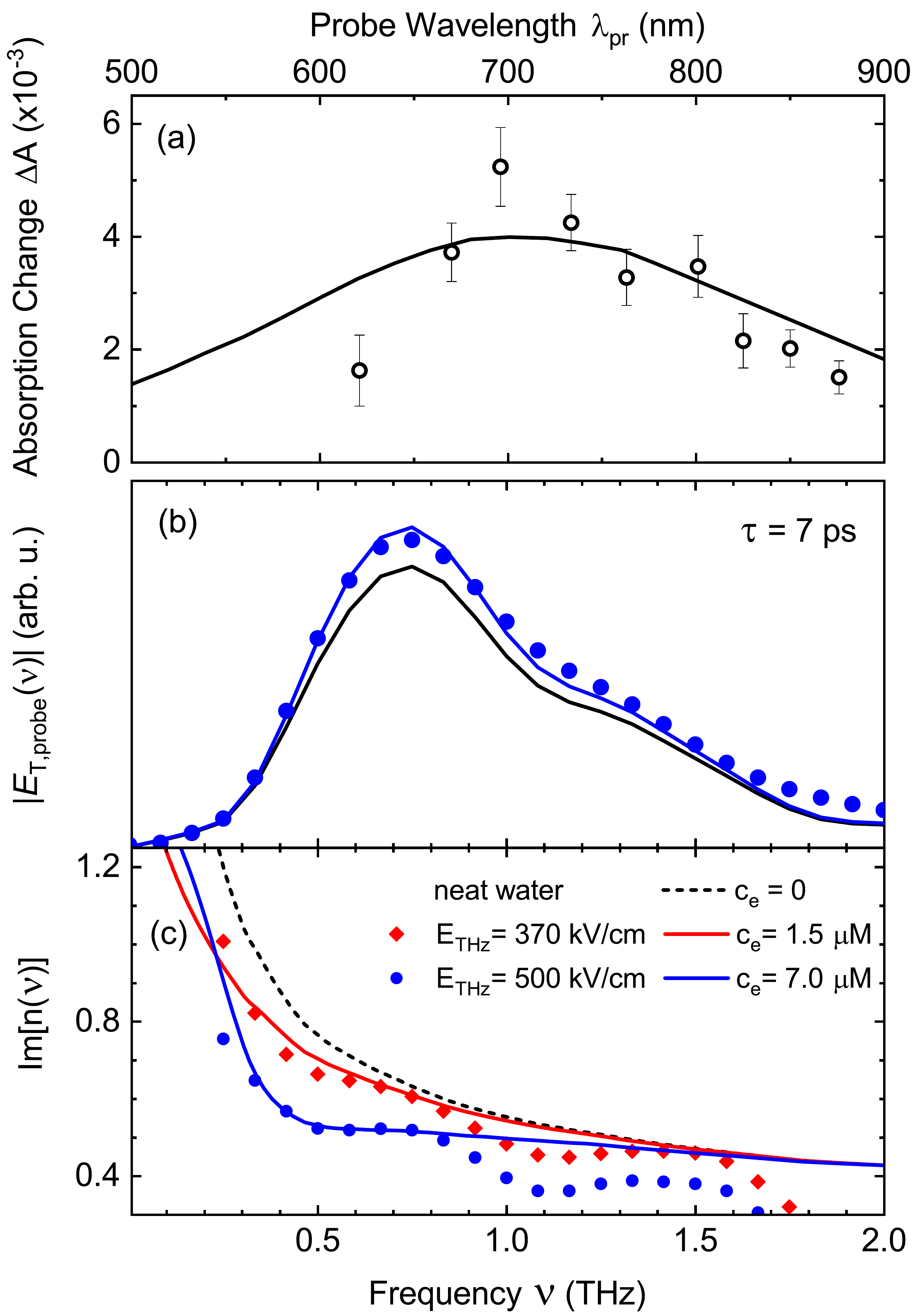}
\caption{(a)~Transient electronic absorption of solvated electrons generated in presence of an external THz field with a transmitted peak amplitude of 1.9 MV/cm. The change of absorbance $\Delta A = -\ln(T/T_0)$ at a delay time of 300 ps is plotted as a function of wavelength (symbols, $T$, $T_0$: sample transmission with and without THz excitation). The solid line represents the (scaled) absorption spectrum reported in Ref. \citenum{Hare2008}. (b) Spectrum of THz probe pulses transmitted after THz excitation of the water sample. The spectrum without THz pumping (solid line) and spectrum after excitation with THz pulses of a peak field strength of 500 kV/cm (symbols, pump-probe delay 7 ps) are shown. (c) Imaginary part of the refractive index Im$(n(\nu))$ from experiment (symbols) and calculation (solid lines). Adapted from ref.~\citenum{Ghalgaoui2020} published under an American Chemical Society Author Choice License.}
\label{fig:data}
\end{figure}

A basic spectroscopic probe for the generation of solvated electrons is their optical absorption. The absorption spectrum of equilibrated electrons solvated in water is plotted as a solid line in Fig. \ref{fig:data}(a). \cite{Hare2008} The symbols in Fig.~\ref{fig:data}(a) represent a transient absorption spectrum of the water sample observed at a time delay of 300 ps after excitation by a THz transient with a peak electric field of 1.9 MV/cm transmitted through the sample. The transient spectrum is close to the equilibrium absorption spectrum of solvated electrons and, thus, gives evidence of electron generation in presence of the THz field. While the THz field is suffcient for inducing a persistent separation of the electron from the parent ion, it is much too small to ionize water molecules directly. Instead, the ionization process is driven by the fluctuating electric field from the water dipoles. The concentration of generated electrons is estimated from the measured absorption change and the known molar extinction coefficient \cite{Hare2008}. For a THz peak field of 1.9 MV/cm (Fig. \ref{fig:data}a), the electron concentration is   $c_e \approx 2 \times 10^{-5}$ M. 

The impact of electron generation on the THz dielectric response of water was studied in 2D-THz experiments. Here, we focus on THz pump/THz probe experiments. The phase-resolved detection of the THz probe pulse transmitted through the water sample allows for separating pump-induced changes of the frequency dependent real part ${\rm Re}[n(\nu)]$ and imaginary part ${\rm Im}[ n(\nu)]$ of the refractive index $n(\nu) = \sqrt{{\rm Re}[\epsilon(\nu)] + i{\rm Im}[\epsilon(\nu)]}$ where $\epsilon(\nu)$ is the dielectric function. The absorption coefficient is given by $\alpha(\nu) =(4 \pi \nu/c_0){\rm Im}[ n(\nu)]$ [$c_0$: vacuum speed of light]. Figure \ref{fig:data}(b) displays spectra $\vert E(\nu) \vert$ of the transmitted THz probe pulses after THz pumping (symbols) and without excitation (black solid line), as derived from time-resolved transients by a Fourier transform. For a delay time of 7 ps after excitation, the transmitted probe field exhibits a larger amplitude than without excitation, demonstrating a decrease of THz absorption. The transient ${\rm Im}[ n(\nu)]$ is plotted in Fig. \ref{fig:data}(c) for two different peak values of the THz pump field (symbols) and displays a broadband decrease compared to neat water (dashed line). 
There is a concomitant change of the real part of the refractive index ${\rm Re}[n(\nu)]$ which has been analyzed in Ref.~\citenum{Ghalgaoui2020}.    

Measurements with THz excitation pulses of different electric field amplitudes reveal a threshold behavior of the nonlinear response. For transmitted pump peak fields below 200 kV/cm, nonlinear changes of ${\rm Re}[n(\nu)]$ and ${\rm Im}[n(\nu)]$ are absent. On the other hand, excitation pulses with a transmitted peak field of 250 kV/cm induce nonlinear changes, as has been discussed in detail in Refs. \citenum{Ghalgaoui2020} and \citenum{Reimann2021}. The occurrence of a threshold excitation field confirms the transport scenario outlined in section 3.2 according to which an electron generated by tunneling ionization needs to acquire a ponderomotive energy $U_p \approx 11$~eV for an irreversible spatial separation from its parent ion. In the experiment, the relevant external field is identical to the transmitted field of the excitation pulse \cite{Elsaesser2019}. Its  threshold value of approximately 250 kV/cm agrees very well with the estimate based on eq. (1).

Electron motions driven by the THz field represent a time-dependent electric current in the water sample. This current gives rise to the emission of an electric field $E_{em}$ which is given by $E_{em}(t)=E_i(t) - E_T(t)$, the difference between the incoming THz field $E_i(t)$ and the total transmitted field $E_T(t)$. The current density $j(t)$ can be derived from the measured $E_{em}(t)$ via $j(t)=2E_{em}(t)/(Z_0d)$. Here, $Z_0 = 1/(\epsilon_0c_0)=377$~$\Omega$ ($\epsilon_0$: vacuum permittivity) and $d$ is the thickness of the water jet which is much smaller than the THz wavelength. For a maximum transmitted driving field $E_T = 500$~kV/cm, the measured emitted field has a maximum value of $E_{em} = 200$ V/cm, giving a maximum current density $j = 2.12$  A/cm$^2$. This corresponds to a maximum velocity of electrons of $v = j/(n\cdot e) = 265$ m/s [$n=5 \times 10^{14}$ electrons/cm$^3$ corresponding to $c_e=5 \times 10^{-6}$~M, $e$: elementary charge]. The resulting distance over which the electrons travel during the THz pulse of 1 -2 ps duration is 260 to 530 \AA, corresponding to many water layers between the generation site and the localization site of the electrons. At such long distances, the attractive electric interaction between the electron and its parent ion is safely screened and, thus, negligible.  

The impact of solvated electrons on the THz dielectric function or refractive index of water has been modelled by a local field or Clausius-Mossotti approach for describing the dielectric response (cf. Sec. 4).\cite{Hannay1983,Ghalgaoui2020} Each solvated electron contributes to the local field in the liquid via its frequency-dependent electric polarizability and, thus, modifies the refractive index. The solid lines in Fig.~\ref{fig:data}(c) represent changes of ${\rm Im}[ n(\nu)]$ calculated from a local field model for the electron concentrations given in the inset. The calculated response is in good agreement with the experimental result. The spectral modulation of the latter originates from the frequency-dependent transmission function of the experimental THz setup.

In summary, spontaneous tunneling ionization of water molecules induced by  the strong fluctuating electric field in water together with electron transport in an external THz field allows for generating solvated electrons. The yield of this generation process sets a benchmark for the amplitudes of the fluctuating electric field on the order 200 to 300 MV/cm and, thus, provides a critical criterion for testing and comparing different theoretical water models.   The scenario of field-induced ionization and charge separation outlined in section 3.2 is fully confirmed by the time-resolved THz experiments and expected to be relevant for a broader range of polar liquids.

\section{Many-body effects and polaron behavior of solvated electrons
\label{sec:vib_exc_dissipation}}

\subsection{Transverse and longitudinal elementary excitations}

Most studies of the dielectric properties of liquid water have considered a frequency dependent dielectric function $\epsilon(\omega)$ ($\omega = 2 \pi \nu$) but neglected any spatial dispersion. Spatial dispersion introduces a $q$-vector dependence of the dielectric response, i.e., $\epsilon({\textbf q}, \omega)$, which accounts for the discrete short-range molecular structure of the liquid and, in particular, allows for distinguishing transverse (T) and longitudinal (L) elementary excitations. In general, the dielectric function is a tensorial quantity \cite{Landau1984,Bopp1998} which can be decomposed in a T- and L-part with respect to the direction of an external electric field ${\mathbf E}(\mathbf{r},t)$:
%
\begin{eqnarray}
\epsilon_{ij}(\mathbf{q},\omega)&=&\epsilon_{\rm T}(q,\omega)\left(\delta_{ij}-\frac{q_i q_j}{q^2}\right) + \epsilon_{\rm L}(q,\omega)\frac{q_i q_j}{q^2} \label{eq:epsTensor}
\end{eqnarray}      
with the transverse $\epsilon_{\rm T}(q,\omega)$ and longitudinal dielectric function $\epsilon_{\rm L}(q,\omega)$  and the spatial directions $i,j=x,y,{\rm or}\; z$. The longitudinal $\epsilon_{\rm L}(q,\omega)$ describes the linear response to plane-wave driving fields with $\mathbf{E}\parallel \mathbf{q}$, whereas $\epsilon_{\rm T}(q,\omega)$ that to fields with $\mathbf{E}\perp \mathbf{q}$. The T- and L-components of $\epsilon_{ij}(\mathbf{q},\omega)$ of liquid water have different imaginary parts in the limit $q\rightarrow 0$, i.e., on a macroscopic length scale.

\begin{figure}[t]
\includegraphics[width=0.75\columnwidth]{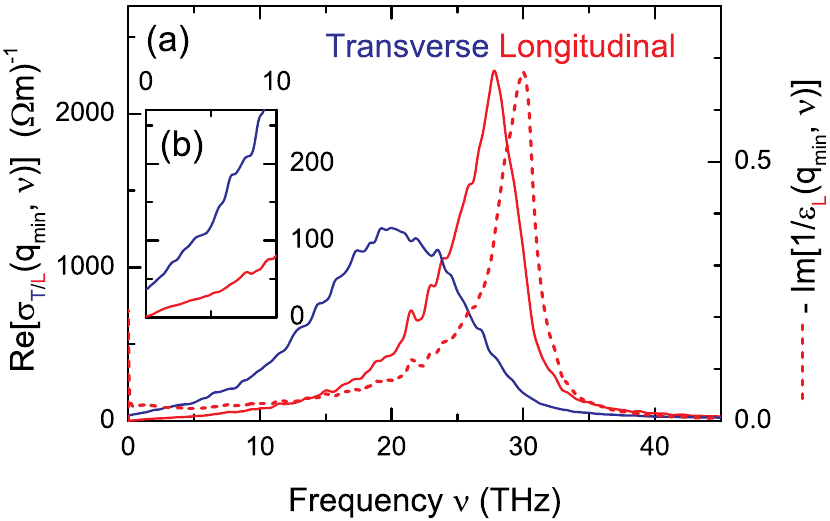}
\caption{(a) Real part of the frequency-dependent conductivity of water ${\rm Re}[\sigma_{\rm T,L}(q_{\rm min},\nu)]$ (solid lines) calculated from a 100-ps long MD trajectory 
(taken from Ref.~\citenum{Ghalgaoui2020})
with the help of the fluctuation-dissipation theorem $[\nu=\omega/(2\pi)]$. Blue:~transverse (T) conductivity as typically measured in the linear optical response. Red:~longitudinal (L) conductivity of water. Longitudinal elementary excitations show resonances in the imaginary part of the inverse dielectric function $-{\rm Im}[1/\epsilon_{\rm LO}(q_{\rm min},\nu)]$ (red dashed line). (b) Low-frequency part of the T and L conductivity spectra shown in panel (a).}
\label{fig:polaron}
\end{figure}

We now derive the T- and L- components of the dielectric response of water from the MD simulation described in section 3 and Ref.~\citenum{Ghalgaoui2020} with the help of the fluctuation-dissipation theorem \cite{Kubo1966}. Instead of the complex dielectric function (eq. \ref{eq:epsTensor}), we consider the complex conductivity tensor $\sigma_{ij}({\mathbf q},\omega)$ which is related to $\epsilon_{ij}(\mathbf{q},\omega)$ by $\epsilon_{ij}(\mathbf{q},\omega)= {\rm Re}[\epsilon_{ij}(\mathbf{q},\omega)] + i\; {\rm Im}[\epsilon_{ij}(\mathbf{q},\omega)]=\delta_{ij}+(1/\omega\epsilon_0){\rm Im}[\sigma_{ij}(\mathbf{q},\omega)]+(i/\omega\epsilon_0){\rm Re}[\sigma_{ij}(\mathbf{q},\omega)]$. The current density $\mathbf{j}(\mathbf{q},\omega)$ is given by 
$\mathbf{j}(\mathbf{q},\omega) = \sigma (\mathbf{q},\omega) \mathbf{E}(\mathbf{q},\omega)$ and in real space connected to the macroscopic polarization by $\mathbf{j}(\mathbf{r},t)=d\mathbf{P}(\mathbf{r},t)/dt$. The conductivity tensor is written as 
\begin{eqnarray}
\sigma_{ij}(\mathbf{q},\omega)&=&\sigma_{\rm T}(q,\omega)\left(\delta_{ij}-\frac{q_i q_j}{q^2}\right) + \sigma_{\rm L}(q,\omega)\frac{q_i q_j}{q^2},\label{eq:condTensor}.
\end{eqnarray}    
Applying the fluctuation-dissipation theorem, the transverse, e.g., $j_x\perp q_y$, and longitudinal conductivity, e.g., $j_x\parallel q_x$, are determined by the current-density-fluctuation-correlation functions defined as follows \cite{Nozieres1999,Mahan2000}:
\begin{eqnarray}
\sigma_{\rm T}(q,\omega)&=&\frac{V_{\rm box}}{k_{\rm B}T}\int_0^{\infty} j_x(0,q,0,0)\;j_x^{\ast}(0,q,0,t)\;e^{i\omega t}\;dt,\label{eq:sigmaTR}\\
\sigma_{\rm L}(q,\omega)&=&\frac{V_{\rm box}}{k_{\rm B}T}\int_0^{\infty} j_x(q,0,0,0)\;j_x^{\ast}(q,0,0,t)\;e^{i\omega t}\;dt.\label{eq:sigmaLO}
\end{eqnarray}

We calculated the $\mathbf{q}$-vector and time-dependent current densities $j_x(q_x,q_y,q_z,t)$, $j_y(q_x,q_y,q_z,t)$, and $j_z(q_x,q_y,q_z,t)$ determined by the time-dependent positions $\mathbf{R}_n^{\rm O}(t)$, $\mathbf{R}_m^{\rm H}(t)$ and velocities $\mathbf{v}_n^{\rm O}(t)$, $\mathbf{v}_m^{\rm H}(t)$ of all oxygen (O: charge $Q_{\rm O}$) and hydrogen atoms (H: charge $Q_{\rm H}$) in the MD box (volume $V_{\rm box}$) with periodic boundary conditions according to
\begin{eqnarray}
\mathbf{j}(q_x,q_y,q_z,t)&=& \frac{Q_{\rm O}}{V_{\rm box}}\sum_n \exp\left[i\mathbf{q}\;\mathbf{R}_n^{\rm O}(t)\right]\mathbf{v}_n^{\rm O}(t) + \frac{Q_{\rm H}}{V_{\rm box}}\sum_m \exp\left[i\mathbf{q}\;\mathbf{R}_m^{\rm H}(t)\right]\mathbf{v}_m^{\rm H}(t).
\end{eqnarray}    

\noindent Results are presented in Fig.~\ref{fig:polaron}(a). The solid lines show the real part of the frequency-dependent conductivity ${\rm Re}[\sigma_{\rm T, L}(q_{\rm min},\nu)]$ of liquid water at room temperature as a function of frequency $\nu=\omega/(2\pi)$. For the cubic box of Ref.~\citenum{Ghalgaoui2020}, the smallest possible wavevector is $|\mathbf{q}_{\rm min}|=2\pi V_{\rm box}^{-1/3} \approx 3\times 10^9$~m$^{-1}$.  

The blue curve in Fig.~\ref{fig:polaron}(a) shows the real part of the transverse conductivity (eq. \ref{eq:sigmaTR}) which is measured in the linear optical response. The MD simulation reproduces the librational absorption band of water in good agreement with experiment \cite{Zelsmann1995}. 
The red solid curve shows the real part ${\rm Re}[\sigma_{\rm L}(q_{\rm min},\nu)]$ of the longitudinal conductivity (eq. \ref{eq:sigmaLO}). This quantity is not accessible in the linear optical response using transverse electromagnetic waves only. The spectrum of ${\rm Re}[\sigma_{\rm L}(q_{\rm min},\nu)]$ differs from its transverse counterpart in several respects: 
\\(i) The longitudinal conductivity resonance (eq. \ref{eq:sigmaLO}) is blue shifted relative to its transverse counterpart (eq. \ref{eq:sigmaTR}) and displays a smaller spectral width.
\\(ii) At low frequencies $\nu \rightarrow 0$ shown in Fig. ~\ref{fig:polaron}(b), the L- and T-components exhibit a markedly different limiting behavior. While the transverse conductivity (blue curve) approaches a finite value, its longitudinal counterpart decays much faster  for $\nu \rightarrow 0$ and vanishes at $\nu = 0$. Since the real part of conductivity is connected with dissipative processes, any elementary longitudinal excitation of liquid water is largely underdamped at frequencies $\nu < 2$~THz.       

\subsection{The solvated electron in a polaron picture}

Both stationary and dynamical properties of the solvated electron can be calculated in the polaron picture, using the  wavevector and frequency-dependent longitudinal dielectric function $\epsilon_{\rm L}(q,\omega)$. The most elementary polaron model is the so-called dielectric continuum model which is similar to the majority of polaron theories developed in solid state physics, e.g., Refs.~\citenum{Lee1953,Frohlich1954,Peeters1985,Gaal2007}, and based on a linear dynamical screening of Coulomb interaction.

The dielectric displacement field $\mathbf{D}(\mathbf{r},t)$ due to a gaussian electron wavepacket
\\\mbox{$\psi(\mathbf{r})=(\eta/\pi)^{3/2}\exp[-\eta\;\mathbf{r}^2]$}, which has a time-dependent width [determined by $\eta(t)$] centered around the time-dependent position $\mathbf{r}_0(t)$, is given by\cite{Bowlan2012}
\begin{eqnarray}
\mathbf{D}(\mathbf{r},t)&=&-\nabla\left[\int\left(\frac{2\eta(t)}{\pi}\right)^{\frac{3}{2}}\exp\left(-2\eta(t)|\mathbf{r}-\mathbf{r}'|^2\right)\frac{e}{|\mathbf{r}_0(t)-\mathbf{r}'|}d^3\mathbf{r}'\right].\label{eq:displace}
\end{eqnarray}    

The semiclassical force field acting back on the electron wavepacket when propagating and shrinking in the medium is given by the electric field distribution $\mathbf{E}(\mathbf{r},t)$ with 
\begin{eqnarray}
\mathbf{E}(\mathbf{q},\omega)&=&\frac{\mathbf{D}(\mathbf{q},\omega)}{\epsilon_0}\left[\frac{1}{\epsilon_{\rm L}(q,\omega)}-\frac{1}{\epsilon_{\rm L}(q,\infty)}\right].\label{eq:electr}
\end{eqnarray}  

$\mathbf{E}(\mathbf{q},\omega)$ and $\mathbf{D}(\mathbf{q},\omega)$ is the four-dimensional Fourier transform of $\mathbf{E}(\mathbf{r},t)$ and $\mathbf{D}(\mathbf{r},t)$, respectively. During the solvation process of the initially free electron, the force-field distribution \eqref{eq:electr} organizes both the deceleration in the medium via a friction force \cite{Bowlan2012} and the concomitant reduction of the wavepacket size down to its final (minimal) value [i.e. $\eta(t) \rightarrow \eta_{\rm se}$] of the solvated electron in thermal equilibrium. The latter value is determined by the balance between the electrostatic energy reduction caused by solvation and the kinetic energy contained in the zero-point motion of the electron wave packet. Similar to the results obtained using Feynman's  path-integral method \cite{Peeters1985} the energy reduction in thermal equilibrium is determined by longitudinal elementary excitations of water with energies $\hbar\omega_{\rm L}>k_{\rm B}T$. The concept presented in  Ref.~\citenum{JingYan1989} gives a solvation energy
\begin{eqnarray}
E_{\rm solv}(\eta_{\rm se})&=& \frac{-1}{4\pi\epsilon_0}\int d^3 \mathbf{q}\left[|\mathbf{D}_{\eta_{\rm se}}(\mathbf{q})|^2-|\mathbf{D}_{\eta = 0}(\mathbf{q})|^2\right]\left[\frac{1}{\epsilon_{\rm L}(q,\infty)}-\frac{1}{\epsilon_{\rm L}(q,k_{\rm B} T/\hbar)}\right]\label{eq:Mukamel}
\end{eqnarray} 
The kinetic energy contained in the zero-point motion of the electron wavepacket is:
\begin{eqnarray}
E_{\rm zpm}(\eta_{\rm se})&=&\frac{3\hbar^2}{4 m_e}\;\eta_{\rm se}\label{eq:zpm}
\end{eqnarray}
corresponding to the total energy minimum defined by $d[E_{\rm zpm}(\eta_{\rm se})+E_{\rm solv}(\eta_{\rm se})]/d\eta_{\rm se}=0$. Thus, the dielectric continuum model provides a ground-state energy of the solvated electron represented by a gaussian wavepacket which is self-consistently determined by the longitudinal dielectric function $\epsilon_{\rm L}(q,\omega)$ of water only. The function $\epsilon_{\rm L}(q,\omega)$ contains contributions from nuclear motions (cf. eq. \ref{eq:sigmaLO}) and from the off-resonant electronic polarizability of water molecules. Since all contributions to polarizability originate from spatially localized dipoles, the Clausius-Mossotti relation \cite{Hannay1983} allows for combining the nuclear [$\epsilon_{\rm L}^{\rm nuc}(q,\omega)$] and electronic contributions [$\epsilon_{\rm L}^{\rm el}(q,\omega)$] to the total longitudinal dielectric function via:    
\begin{eqnarray}
3\;\frac{\epsilon_{\rm L}(q,\omega)-1}{\epsilon_{\rm L}(q,\omega)+2}&=& 3\;\frac{\epsilon_{\rm L}^{\rm nuc}(q,\omega)-1}{\epsilon_{\rm L}^{\rm nuc}(q,\omega)+2} + 3\;\frac{\epsilon_{\rm L}^{\rm el}(q,\omega)-1}{\epsilon_{\rm L}^{\rm el}(q,\omega)+2}  \label{eq:nucplusel}
\end{eqnarray}            
In the simplest approximation, the off-resonant electronic contribution does not depend on the wave vector $q$ and is determined by the refractive index of water above all vibrational resonances, e.g., $\epsilon_{\rm L}^{\rm el}(q,\omega)=n^2_{\rm water}(500\;{\rm THz})$.

\begin{figure}[t]
\includegraphics[width=0.8\columnwidth]{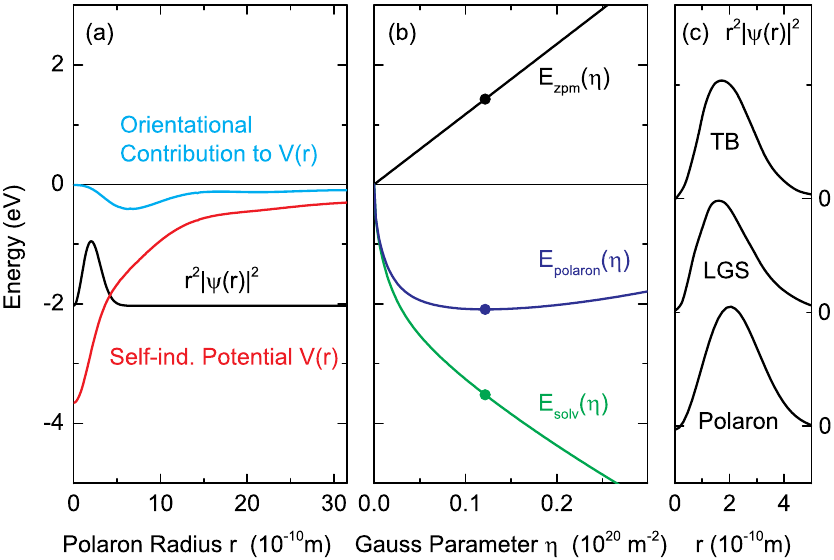}
\caption{(a)~Self-induced potential $V(r)$ of the solvated electron (red line) and radial density $r^2|\Psi(r)|^2$ (black line) in the polaron model. Cyan line: contribution to $V(r)$ caused by the low-frequency orientational alignment of water molecules in the solvation shell. (b)~Ground-state energy of the polaron (blue line) as a function of the gauss parameter $\eta$ in eqs. \eqref{eq:Mukamel} (green line) and \eqref{eq:zpm} (black line). The minimum of $E_{\rm polaron}(\eta)$ at $\eta_{\rm se}$ is indicated by symbols. (c)~Comparison of the radial densities $r^2|\Psi(r)|^2$ in the Turi-Borgis (TB), Larsen-Glover-Schwartz (LGS), and the polaron model.}
\label{fig:polpot}
\end{figure}

We now apply the polaron approach to calculate the self-consistent potential of the electron electron, using the longitudinal dielectric function $\epsilon_{\rm L}^{\rm nuc}(q,\omega)$ from Ref.~\citenum{Bopp1998}. The latter has been derived from classical MD simulations of water, including the high-frequency intramolecular O-H bending and O-H stretching modes. The electronic polarizability 
 $\epsilon_{\rm L}^{\rm el}(q,\omega)=n^2_{\rm water}(500\;{\rm THz})$ is added according to eq.~\eqref{eq:nucplusel}. The results are summarized in Fig.~\ref{fig:polpot}. In panel (a), the red curve shows the self-induced potential $V(r)$ as a function of the polaron radius $r$. The polaron model predicts a ground-state energy of the solvated electron which is $2.03$~eV below the continuum states as determined by minimizing $E_{\rm polaron}(\eta) = E_{\rm zpm}(\eta) + E_{\rm solv}(\eta)$ through a variation of the gauss parameter $\eta$ in balance of the  solvation energy [eq. \ref{eq:Mukamel}, green curve in panel (b)] and the kinetic energy in the zero-point motion [eq. \ref{eq:zpm}, black curve]. The corresponding radial density $r^2|\Psi(r)|^2$ is shown as black line in panel (a). In Fig.~\ref{fig:polpot}(c) we compare the radial densities $r^2|\Psi(r)|^2$ predicted by the Turi-Borgis (TB), Larsen-Glover-Schwartz (LGS), and the polaron model. Interestingly, the three models give almost identical electron wave packets.  

The polaron model allows for analyzing the different contributions to its self-induced potential (red curve in Fig.~\ref{fig:polpot}a). Such analysis shows that the electronic polarizability of water molecules, their intra-molecular vibrations and the librational contribution to the polarizability are mainly responsible for the size of the electron wavepacket. In contrast, contributions to $V(r)$ caused by the orientational alignment of water molecules in the solvation shell which occurs at frequencies below some 10 THz, have no influence on the wavepacket size, but on the depth and detailed shape of the self-induced potential (cyan curve in Fig.~\ref{fig:polpot}a). Obviously, this contribution is predominantly located at radial distances outside the electron wavepacket.  

The calculated depth of the electron potential depends on the particular water model applied for calculating the dielectric response. The MD simulations of Ref.~\citenum{Bopp1998} are based on a entirely classical nonpolarizable water model, giving a smaller potential depth than the cavity model.  Moreover, it should be emphasized that the polaron concept presented here is based on {\em linear} dynamical screening of the electron's displacement field (eq. \ref{eq:displace}) according to eq. \eqref{eq:electr} and {\em does not} account for major changes of solvent structure around the electron, e.g., the formation of cavities and others. A description of the latter effects requires a {\em nonlinear} dynamical screening concept.
Models for $\epsilon_{\rm L}(q,\omega)$ which address such shortcomings, e.g., by  a quantum treatment of the O-H bending and stretching vibrations and introduction of polarizable water molecules, are expected to translate into a deeper electron potential with, however, a similar spatial extension of the electron wavepacket. Such work is beyond the scope of the present article.

The dielectric continuum model \cite{JingYan1989,Bowlan2012} intrinsically contains the dynamic response of the combined system, i.e., the electron plus its solvation shell, via the linear relation \eqref{eq:electr}. In particular, the model allows for including modifications of $\epsilon_{\rm L}(q,\omega)$ caused by the presence of photogenerated electrons. Due to their localized character when solvated, one can use the Clausius-Mossotti relation to calculate such modifications, as will be shown in the next section.

\subsection{Polaron oscillations of electrons solvated in water}
     
\label{sec:PolOsc}     
     
\begin{figure}[t]
\includegraphics[width=0.5\columnwidth]{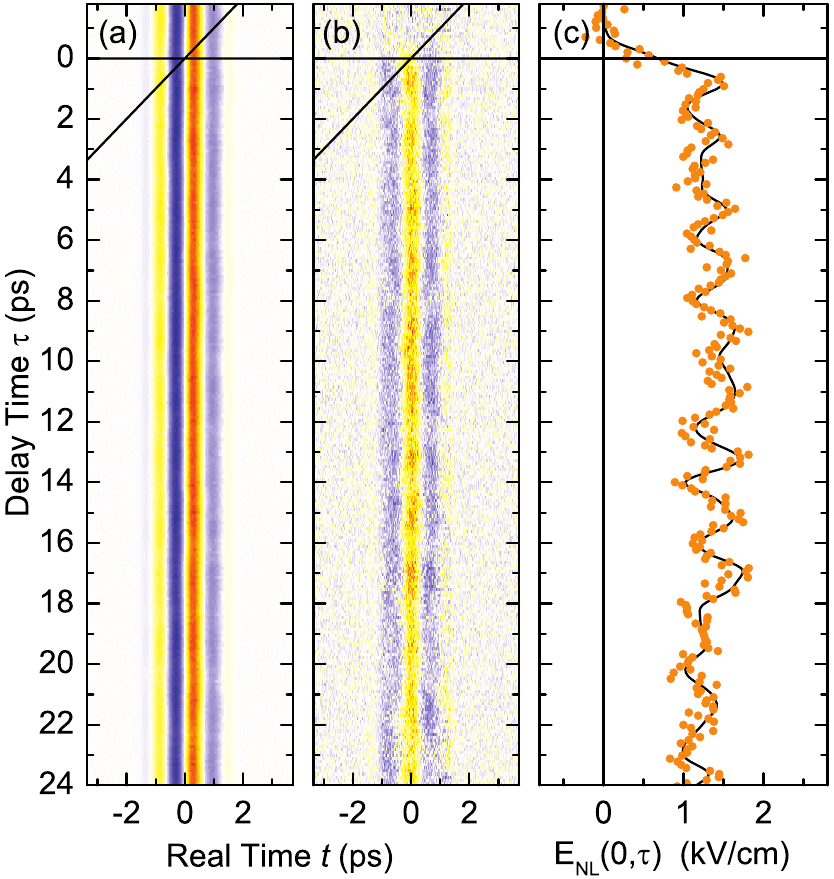}
\caption{Nonlinear THz response of liquid water after femtosecond generation of free electrons (electron concentration of $c_e=16~\mu$M) by an intense 800-nm pulse indicated by the tilted black line. (a)~Contour plot of the two-dimensional scan along real time $t$ and pump-probe delay $\tau$ of the THz-probe pulse $E^0_\text{pr}(t)$ (from blue: $-140$~kV/cm to red: $+140$~kV/cm) transmitted through the unexcited sample, and (b)~the nonlinear signal field $E_\text{NL}(t,\tau)$. (c)~Orange symbols: nonlinear signal $E_\text{NL}(0,\tau)$ as a function of  $\tau$ averaged over real times in the interval $-70\;{\rm fs}<t<+70\;{\rm fs}$. Black line: 2D-Fourier filtered nonlinear signal.}
\label{fig:2Dpolaron}
\end{figure}

\begin{figure}[t]
\includegraphics[width=0.9\columnwidth]{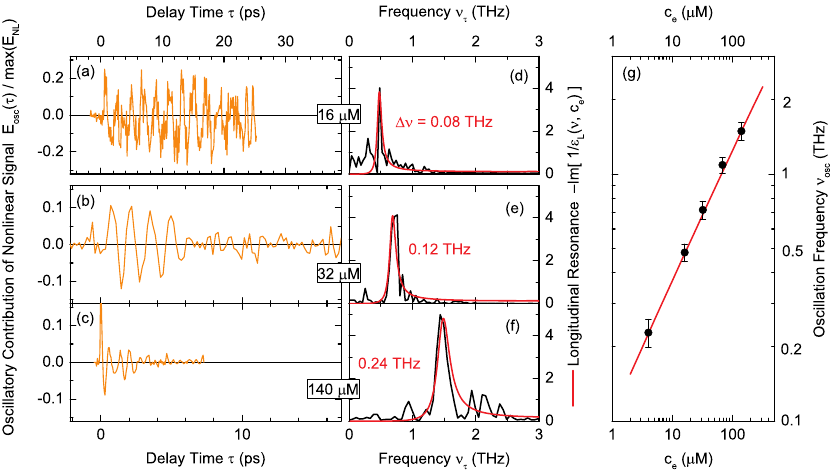}
\caption{(a)--(c)~Oscillatory contributions $E_\text{osc}(\tau)$ as a function of pump-probe delay $\tau$ for different concentrations $c_e$ of hydrated electrons.  
\mbox{(d)--(f)}~Black curves: Fourier transforms of the transients
shown in panels (a) to (c). Red curves: Spectra $-\text{Im}[1/\epsilon_{\rm L}(\nu,c_e)]$ calculated from Eq.~\eqref{eq:CM}  with $\gamma=0$ together with their spectral widths $\Delta\nu$ (FWHM) for electron concentrations $c_e$ as indicated. (g)~Oscillation frequency as a function of $c_e$. Adapted from ref.~\citenum{Ghalgaoui2021} published by the American Physical Society under the Creative Commons Attribution 4.0 International License.}
\label{fig:Oscpolaron}
\end{figure}

Polarons subject to the fluctuating electric field of the water environment represent an electron wavepacket fluctuating collectively with longitudinal elementary excitations in its vicinity. On top of the longitudinal excitations of neat water discussed so far  (cf. red dashed line in Fig.~\ref{fig:polaron}a), there are specific polaronic degrees of freedom of a longitudinal character. Ultrafast electron acceleration or deceleration should impulsively excite such longitudinal modes, in analogy to the experimentally observed longitudinal-optical (LO) phonon oscillations of polarons in the semiconductor GaAs \cite{Gaal2007,Woerner2010}. The transition of electrons generated in continuum states of water by photoexcitation and/or tunneling ionization into a localized ground state represents a primary deceleration process occurring on a subpicosecond time scale, much shorter than the period of low-frequency longitudinal excitations. Very recently, coherent polaronic oscillations induced by electron deceleration have been observed in femtosecond pump-probe experiments.\cite{Ghalgaoui2021}  The main results are briefly discussed in the following. 

Free electrons are generated in liquid water by multiphoton ionization of water molecules with a femtosecond pulse centered at a wavelength of 800 nm. The dielectric response is probed in transmission by THz pulses which are detected in a phase-resolved way by electrooptic sampling.  The nonlinear signal field is defined as $E_\text{NL}(t,\tau)=E_\text{pr}^\text{pumped}(t,\tau)-E_\text{pr}^0(t)$ where $E_\text{pr}^\text{pumped}(t,\tau)$ is the THz probe field transmitted through the excited sample and $E_\text{pr}^0(t)$ the field transmitted without pump ($\tau$: delay time between the two pulses, $t$: real time as defined by electrooptic sampling). 

Time resolved signal transients recorded with an electron concentration $c_e=16~\mu$M are shown in Fig.~\ref{fig:2Dpolaron}. Panel (a) displays a contour plot (electric-field strength from blue: $-140$~kV/cm to red: $+140$~kV/cm) of the two-dimensional scan of $E_\text{pr}^0(t)$ along real time $t$ and pump-probe delay $\tau$ of the THz-probe pulse $E_\text{pr}(t)$. The temporal position of the 800-nm pump pulse is indicated by the tilted black line. 
Panel (b) shows the nonlinear signal field $E_\text{NL}(t,\tau)$ which displays oscillations as a function of the pump-probe delay $\tau$, most pronounced for the real time $t\approx 0$. In panel (c), the orange symbols represent the nonlinear signal $E_\text{NL}(0,\tau)$ averaged over the  real time interval $-70\;{\rm fs}<t<+70\;{\rm fs}$ as a function of pump-probe delay $\tau$. For visualizing the oscillations best, we show the Fourier-filtered nonlinear signal as a black line.
 The oscillatory part of the signal is isolated by subtracting the step-like contribution and plotted for various electron concentrations $c_e$ in Fig.~\ref{fig:Oscpolaron}.
In panels (a) to (c), the oscillations $E_\text{osc}(\tau)$ are shown in the time domain as a function of pump-probe delay $\tau$. The black lines in panels (d) to (f) give the corresponding Fourier spectra. The oscillation frequency scales with electron concentration $c_e$ as shown in Fig.~\ref{fig:Oscpolaron}(g). 

 A quantitative analysis of this data set needs to include the impact of local electric fields on the individual charges and dipoles. The Clausius-Mossotti relation allows for description of local field effects on the basis of the  following concentration-dependent dielectric function $\epsilon(\nu,c_e)$:    
\begin{eqnarray}
3\frac{\epsilon(\nu,c_e)-1}{\epsilon_{\rm}(\nu,c_e)+2}&=& 3\frac{\epsilon_\text{water}(\nu)-1}{\epsilon_\text{water}(\nu)+2}+ c_e\,N_A\alpha_\text{el}(\nu),\label{eq:CM}\\
\text{with}\qquad\alpha_\text{el}(\nu) &=& -\frac{e^2}{\epsilon_0 m_e \left[ (2\pi\nu)^2+i\gamma(2\pi\nu)\right]}\nonumber
\end{eqnarray}            
Here, electrons with a polarizability $\alpha_\text{el}(\nu)$ are added to the neat water dielectric response [$\epsilon_\text{water}(\nu)$],
with the Avogadro constant $N_A$, elementary charge $e$, electron mass $m_e$, and local friction rate $\gamma$. Since the longitudinal excitations of the electron wave packet, e.g., size oscillations, do not experience much damping beyond the one contained in $\epsilon_\text{water}(\nu)$, $\gamma=0$ is used. For wavevectors $q < \sqrt{4 \eta}$, i.e., the range covered by the size of the electron wavepacket, the longitudinal dielectric function displays a weak $q$-dependence only which is neglected here. 

Equation \ref{eq:CM} contains the longitudinal dielectric functions (cf. eq. \ref{eq:nucplusel}), which, however, are not directly accessible in optical and dielectric measurements and, thus, barely characterized. For the present analysis, we instead use the well-characterized transverse dielectric function $\epsilon_{\rm water}(\nu)$ of water \cite{Zelsmann1995}. In the THz frequency range, the real parts of the transverse and longitudinal dielectric functions are similar, while the imaginary part of the longitudinal function is substantially smaller \cite{Bopp1998}. The imaginary part of the transverse $\epsilon_{\rm water}(\nu)$ is broadened by the pronounced damping of transverse excitations in the liquid. As a result, replacing the longitudinal by the transverse $\epsilon_{\rm water}(\nu)$ in the calculations tends to overestimate the spectral width of the polaron resonances.

The red lines in Fig.~~\ref{fig:Oscpolaron}(d) to (f) repesent spectra $-\text{Im}[1/\epsilon_{\rm L}(\nu,c_e)]$ calculated from eq.~\eqref{eq:CM} together with their spectral widths $\Delta\nu$ (FWHM) for electron concentrations $c_e$. In panel (g) the calculated resonance frequencies are shown  as a solid line together with the experimental frequencies (symbols) as a function of the electron concentration $c_e$. In all cases, the calculated spectra and frequency positions are in excellent agreement with experiment.   
The good agreement demonstrates that the oscillatory signals are due to impulsively excited coherent longitudinal oscillations of the polaron. As expected from the extremely small damping of the longitudinal conductivity (eq. \ref{eq:sigmaLO}) at frequencies $\nu < 2$~THz,  the polaron oscillations are highly underdamped and  last for up to 20~ps and beyond at low electron concentrations $c_e$.

In a system of charged and/or polar particles, the time scale of damping of longitudinal excitations is set by the charge-density-fluctuation-correlation function, which is, via the continuity equation of electric charge, connected to the longitudinal current-density-fluctuation-correlation function (eq. \ref{eq:sigmaLO}).\cite{Bopp1998,Nozieres1999,Mahan2000,Elton2016} In contrast, the dephasing of transverse excitations is governed by the transverse current-density-fluctuation-correlation function (eq. \ref{eq:sigmaTR}). The latter is much more susceptible to fluctuations and scattering processes than the correlation function of charge density. For instance, elastic scattering may decrease the electric current, but does not change the charge density. Accordingly, transverse excitations are much more strongly damped than longitudinal excitations. 
 
\section{Conclusions}
The results discussed in this article demonstrate the multifaceted character of Coulomb interactions between electrons and their aqueous environment. The ultrafast localization of solvated electrons in a self-consistent potential of some 5 eV depth is mediated by electric forces the electron exerts on the dipolar water molecules in its neighborhood. The stochastic motions of water dipoles at ambient temperature generate a fluctuating electric field which reaches local peak values of several hundred megavolts/cm. As a result, spontaneous tunneling ionization of water molecules arises. While recombination of the released electron and its parent ion prevails under equilibrium conditions, application of an external electric field in the terahertz range allows for separating the charges and, thus, generate persistent solvated electrons. The pronounced polaronic properties of solvated electrons are another manifestation of electric interactions at the molecular level. The coupling of electron motions to longitudinal librational excitations of the liquid results in long-lasting coherent polaron oscillations with a frequency determined by electron concentration. The slow damping of the oscillatory response points to a weak coupling of such longitudinal modes to other low-frequency degrees of freedom and is in marked contrast to the ultrafast decay of transverse excitations.     

The concept of electron generation by tunneling ionization may be extended beyond water, for instance to alcohols and other polar liquids. Another interesting class of systems are individual nucleic acids and/or larger DNA and RNA strands in an aqueous environment. Here, tunneling ionization of nucleic acids and backbone units may arise on top of water ionization, with a strong impact on the structural and chemical properties and the hydration geometries. From a theory point of view, simulations of the ionization and charge separation processes are required to develop a realistic picture of nonequilibrium dynamics at the molecular level. In a similar way, experimental and theoretical studies of the polaronic response of solvated electrons need to cover a broader range of systems and address the different properties of transverse and longitudinal excitations of the molecular ensemble.

\begin{acknowledgement}
This research has received funding from the European Research Council (ERC) under the European Union’s Horizon 2020 research and innovation program (grant agreements No. 833365 and No. 802817). 
\end{acknowledgement}


\bibliography{solvelectron}

\newpage
\noindent TOC Graphic
\begin{figure}
\includegraphics{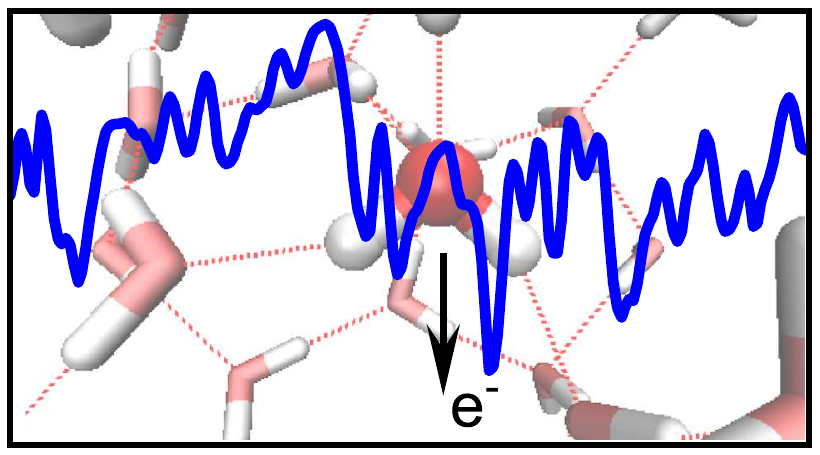}
\end{figure}
\vspace{5mm}

\noindent Author biographies

\singlespacing
\noindent Michael Woerner is a department head at the Max-Born-Institute, Berlin, Germany, and holds a lecturer qualification (Habilitation) in physics at Humboldt University, Berlin. He received a Dr. rer. nat. degree from the Technical University of Munich in 1991 and worked there as a postdoc until 1993. He then joined the Max-Born-Institute in 1993 and spent a postdoc period at Bell Laboratories (Lucent Technologies), Holmdel, in 1997. In 2019, Michael was on sabbatical at ETH Zurich, Switzerland. Michael's  research focuses on ultrafast phenomena in solids and nanostructures with pioneering work in multi-dimensional spectroscopies in the THz frequency range and in femtosecond x-ray diffraction using laser-driven hard x-ray sources.
\\[2mm]
Benjamin Fingerhut is heading the Biomolecular Dynamics theory group at the Max-Born-Institute. He received a Dr. rer. nat. degree from the Ludwig-Maximilians-Universität München (LMU) in 2011 and afterwards joined as a postdoctoral fellow the group of Prof. Shaul Mukamel at the University of California, Irvine (UCI). He joined the Max-Born-Institute in 2014. Benjamin's research focuses on ultrafast phenomena at biological interfaces and the development of efficient numerical methods for the description of condensed phase dissipative quantum dynamics. Benjamin's research is supported by a starting grant of the European Research Council, and he is recipient of the Robin Hochstrasser Young Investigator Award and the Coblentz Award.
\\[2mm]
Thomas Elsaesser is a director at the Max-Born-Institute, and a full professor for experimental physics at Humboldt University, Berlin. He received a Dr. rer. nat. degree from the Technical University of Munich in 1986 and worked there as a research associate until 1993. He spent a postdoc period at AT\&T Bell Laboratories, Holmdel, in 1990 and joined the newly established Max-Born-Institute in 1993. His research focuses on ultrafast phenomena in condensed matter, in particular molecular liquids, biomolecules in their aqueous environment, and inorganic solids. Methods of ultrafast spectroscopy and structure research are combined in his experimental work. Thomas is a fellow of the American Physical Society and the Optical Society of America and has received numerous scientific awards.

\end{document}